\documentclass[aps,prb,twocolumn]{revtex4-2}
\usepackage{graphicx}
\usepackage{chngpage}
\usepackage{bm}
\usepackage{amsmath}
\usepackage{amssymb}
\usepackage{float}
\usepackage{xcolor}
\usepackage[title]{appendix}
\usepackage[english]{babel}
\definecolor{darkgreen}{rgb}{0.0, 0.5, 0.0}

\begin{document}

%\title{Towards a cold source of atomic Tritium}
\title{Cryogenic source of atomic tritium for neutrino-mass measurements and precision spectroscopy}

\author{A. Semakin}
\affiliation{Department of Physics and Astronomy, University of Turku, 20014, Turku, Finland}
\author{J. Ahokas}
\affiliation{Department of Physics and Astronomy, University of Turku, 20014, Turku, Finland}
\author{T. Kiilerich}
\affiliation{Department of Physics and Astronomy, University of Turku, 20014, Turku, Finland}
\author{S. Vasiliev}
\affiliation{Department of Physics and Astronomy, University of Turku, 20014, Turku, Finland}
\author{F. Nez}
\affiliation{Laboratoire Kastler Brossel,Sorbonne Université, CNRS, ENS-PSL Université, Collège de France, Paris, France}
\author{P. Yzombard}
\affiliation{Laboratoire Kastler Brossel,Sorbonne Université, CNRS, ENS-PSL Université, Collège de France, Paris, France}
\author{V.V. Nesvizhevsky}
\affiliation{Institut Max von Laue - Paul Langevin, 71 avenue des Martyrs, Grenoble, France}
\author{E. Widmann}
\affiliation{Marietta Blau Institute for Particle Physics, Austrian Academy of Sciences, Vienna, Austria}
\author{P. Crivelli}
\affiliation{Institute for Particle Physics and Astrophysics, ETH, Zurich, Switzerland}
\author{C. Rodenbeck}
\affiliation{Institute for Astroparticle Physics (IAP), Karlsruhe Institute
of Technology (KIT), Eggenstein-Leopoldshafen, Germany.}
\author{M. Röllig}
\affiliation{Institute for Astroparticle Physics (IAP), Karlsruhe Institute
of Technology (KIT), Eggenstein-Leopoldshafen, Germany.}
\author{M. Schlösser}
\affiliation{Institute for Astroparticle Physics (IAP), Karlsruhe Institute
of Technology (KIT), Eggenstein-Leopoldshafen, Germany.}

\begin{abstract}

We propose a concept for a cryogenic source of atomic tritium at sub-Kelvin temperatures and energies suitable for magnetic trapping. The source is based on the dissociation of solid molecular T$_2$ films below 1 K by electrons from a pulsed RF discharge, a technique recently demonstrated for atomic hydrogen, combined with buffer-gas cooling and magnetic confinement. We analyze the key processes limiting the source performance, adsorption, spin exchange and recombination, and show that atomic tritium fluxes exceeding $10^{15}$~s$^{-1}$ at temperature of $\sim$ 200~mK can be achieved. Such a source would enable Doppler-free two-photon 1S–2S spectroscopy in atomic tritium for high-precision measurements of the triton charge radius, providing a crucial benchmark for bound-state QED and improving the comparison between electronic, muonic, and scattering determinations of nuclear sizes in light systems. Beyond spectroscopy, using atomic tritium source avoids molecular final-state broadening in $\beta$- decay and is therefore necessary for next-generation neutrino-mass measurements; combined with detector technologies such as sub-eV resolution quantum sensors or cyclotron radiation emission spectroscopy, it enables an order-of-magnitude improvement compared to the current best experimental limit.
Additionally, the source can be used to generate a beam of low-field-seeking deuterium atoms for loading magnetic traps, an important benchmark before trapping tritium atoms, which is useful for precision spectroscopy.
\end{abstract}

\maketitle
\section{Introduction}

Studies of atomic hydrogen have played a central role in the development of modern physics. Due to its simplicity, many of its properties can be calculated from the first principles with extremely high accuracy, providing stringent tests of bound-state quantum electrodynamics (QED)~\cite{Jentschura:2022xuc}. The experimental possibility of reaching Bose–Einstein condensation stimulated extensive research on ultracold hydrogen atoms confined by superfluid helium films~\cite{Silvera1980} or magnetic traps~\cite{greytak2000}, where BEC was finally achieved in 1998~\cite{Fried1998}. Renewed interest in such systems has recently emerged in connection with precision spectroscopy~\cite{Brandt2022, PhysRevLett.132.113001,amit2025}, gravitational quantum states~\cite{vasiliev2019, Killian2024}, and comparisons with antihydrogen to test the equivalence principle \cite{Ahmadi:2018eca,Azevedo:2023kgf,Baker:2025ehs}. Extending these studies to the heavier isotopes, deuterium and tritium, remains a major challenge, because of their stronger adsorption on helium surfaces and faster recombination. In the case of tritium, precision spectroscopy of the transition $1\mathrm{S}$–$2\mathrm{S}$ would allow an accurate determination of the triton charge radius, providing an essential benchmark for few-body QED and consistency between the radii obtained from electronic, muonic, and scattering data~\cite{Schmidt_2018}. Such measurements would also contribute to the resolution of persistent discrepancies among charge radii of light nuclei, which remain despite of recent convergence of some proton-radius determinations~\cite{Gao_2022, Maisenbacher2026}.

Tritium is of particular interest because it is radioactive and decays by $\beta$-emission into $^3\mathrm{He}$, an electron, and an antineutrino. Accurate measurements of the electron energy near the endpoint provide a direct determination of the effective electron-neutrino mass. This approach is pursued by several international collaborations, including the Karlsruhe Tritium Neutrino Experiment (KATRIN)~\cite{Aker_2021}, Project~8~\cite{Esfahani_2017}, QTNM~\cite{Amad_2025} and PTOLEMY~\cite{PtolemyJCAP}. The KATRIN experiment, which employs a molecular source of T$_2$ at 80~K, has recently established an upper limit of 0.45~eV for the neutrino mass, with a projected sensitivity of 0.3~eV (both at 90$\%$ confidence level) in the final stage of its current measurement campaign~\cite{katrin2025}. These values remain above the expected lower bound of about 50~meV (9~meV) in the case of an inverted mass hierarchy (normal hierarchy). The normal and inverted hierarchies denote the two possible orderings of the neutrino mass eigenstates $m_1$, $m_2$, and $m_3$ consistent with oscillation data. In the normal (inverted) ordering one has $m_1 < m_2 < m_3$  ($m_3 < m_1 < m_2$), implying different minimal values for the effective electron-neutrino mass $m^2(\nu_e)= \sum_i m_i^2 |U_{ei}|^2$\cite{Otten2008} accessible to direct kinematic searches~\cite{Nufit2024}.
One of the main limitations of the molecular approach is that part of the $\beta$-decay energy is distributed into the ro-vibrational excitations of the final state of T$_2$, which broadens the endpoint spectrum~\cite{Bodine2015}. This has motivated the development of sources of \emph{atomic} tritium, where such molecular effects are absent, offering the possibility of reaching much higher precision in future direct neutrino-mass measurements.

Techniques realized for magnetic trapping of H are now considered for holding large amounts of atomic D and T at temperatures $\lesssim$~100~mK.  Loading such a magnetic trap requires an efficient source of atoms with energies low enough for trapping with existing superconductive magnet technologies. The dissociation of molecular H$_2$, D$_2$ and T$_2$ can be carried out in RF discharge~\cite{Walraven1982, Jochemsen1982}, by thermal cracking \cite{Tschersich2008} or with a pulsed DC discharge combined with supersonic expansion~\cite{Scheidegger2022}. Although large fluxes and high degree of dissociation can be achieved with these methods, the resulting atoms have very large energies and reducing these to the trappable values is a very difficult task, though work is underway on rendering previously successful cooling and slowing methods from H and D sources \cite{Esfahani2025, LindmanThesis} compatible with tritium.

In this work, we propose a concept of a cryogenic source of atomic tritium. Dissociating T$_2$ will be done below 1~K in a thin solid layer inside a dissociator chamber in a pulsed RF discharge similar to what is used to dissociate H$_2$~\cite{semakin2025}. Electrons generated in tritium $\beta$-decay at a mean energy of 5.7~keV propagate inside the solid layer and thus provide additional dissociation increasing the total atomic flux. For the H source, following dissociation, the thermalization of the atoms usually takes place through collisions with the cold walls. However, this will not be feasible for tritium. Therefore, buffer gas cooling with $^4$He or $^3$He vapor combined with radial magnetic confinement is proposed in this work. This method was utilized to cool and trap several atomic and molecular species after laser ablation from the solid target in J.~Doyle's group (see~\cite{Carvalho1999, Hutzler2012} for a review). We evaluate that atomic fluxes of the order of 10$^{15}$ T atoms~s$^{-1}$ can be obtained at gas temperature of $\approx0.4$ (0.2)~K using $^4$He ($^3$He) inside the gas transport tube.

\section{Properties of H, D and T}
\begin{figure}
    \includegraphics[width=\linewidth]{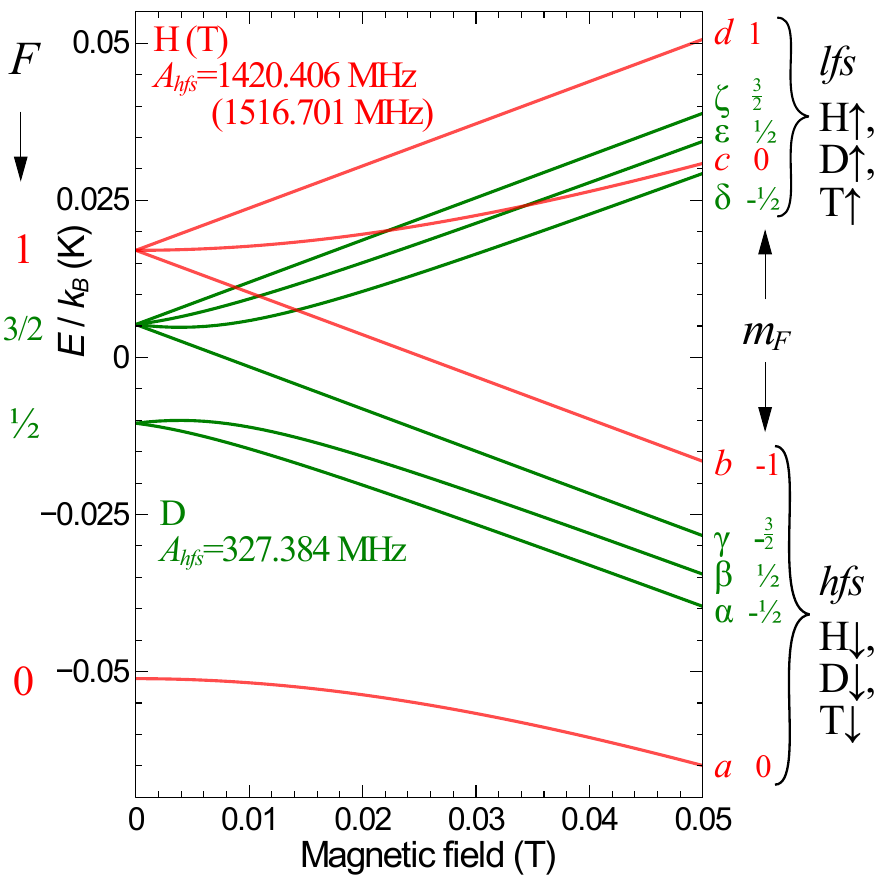}
    \caption{Hyperfine level diagram for H (red) and D (green) and spin states in the low field basis. For T, the diagram is nearly identical to that of H except for the slightly larger value of the hyperfine constant. } 
    \label{HF diagram}
\end{figure}

Magnetically trapped atoms are not interacting with the physical walls. However, the operating principles of the source, which we suggest to use for loading magnetic traps, rely on the physics of interactions of atoms with superfluid helium films. Therefore, first we provide a brief review of the properties of H isotopes essential for our proposal: interactions of atoms with superfluid helium film, surface recombination and relaxation, effects of collisions of atoms with each other and helium vapor. A summary of the relevant data is presented in the Tables~\ref{tab:adsorption_solvation} and ~\ref{cross lengths}. Depending on the type of experiments in a magnetic field, we will follow common notation of high- and low-field seeking hyperfine states (\textit{hfs} and \textit{lfs}), or spin-down (H$\downarrow$) and spin-up (H$\uparrow$) with respect to a static magnetic field directed up, see the hyperfine level diagram in Fig.~\ref{HF diagram}.

\begin{table*}[ht]
    \centering
    \begin{tabular}{||l||c|c|c|c||}
        \hline \hline
        Isotope & $E_{a}$ on $^4$He                                                        & E$_{a}$ on $^3$He          &$E_s$ in $^4$He                             & $l_{\mathrm{rec}} $, \AA \\
        \hline \hline
        H theor & 0.85~\cite{Stwalley1982}, 1.00~\cite{Goodman1989}       & 0.36~\cite{Stwalley1982}   & 36~\cite{Kurten1985, Saarela1993}          &  \\
        \hline
        H exp   & 1.14(1)~\cite{Safonov2001}                                               & 0.39(1)~\cite{Safonov2001} &                                            & 0.4(1)~\cite{Safonov2001} \\
        \hline
        D theor & 2.2~\cite{Stwalley1982}, 2.54~\cite{Goodman1989}  & 1.2~\cite{Stwalley1982}    & 14~\cite{Kurten1985}, 15\cite{Saarela1993} &  \\
        \hline
        D exp   & 3.1(2)~\cite{Mosk2001}                                                   &                            & 13.6(6)~\cite{Reynolds1991}                & $<$30~\cite{Hayden1995}, $>$300~\cite{Mayer1985} \\
        \hline
        T theor & 3.2~\cite{Stwalley1982}, 4.77~\cite{Goodman1989}        & 1.9~\cite{Stwalley1982}    & 6~\cite{Kurten1985}, 7.2~\cite{Saarela1993} &  \\
        \hline
        T exp   &                                                                          &                            &                                             & $>$10000~\cite{Tjukanov1988} \\
        \hline \hline
    \end{tabular}
    \caption{A literature compilation of the data for adsorption $E_{a}$ and solvation $E_s$ energies (in K) along with the recombination cross-length $l_{\mathrm{rec}} $ for different hydrogen isotopes.}
    \label{tab:adsorption_solvation}
\end{table*}

\subsection{Interaction with walls covered by a superfluid helium film}
Finding surfaces that would be most suitable for the confinement of atomic hydrogen was a key issue at the early stage of experimental work with cold H$\downarrow$ gas. Liquid helium, due to its smallest density and the formation of a superfluid film below 1~K, turned out to be the only choice to stabilize H$\downarrow$ and D$\downarrow$~\cite{Silvera1980, Silvera1980D}. The adsorption energy $E_a=1.14(1)$~K~\cite{Safonov2001} of H on pure $^4$He and $E_a\approx 0.36$~K~\cite{Safonov2001} $^3$He-$^4$He has the lowest value of any other surface. 

The density of the adsorbed gas increases exponentially with decreasing temperature as $\exp (E_a/T)$, which leads to a very fast surface recombination, limiting the accessible experimental range from below at around  $0.1 E_a$. Due to a larger mass, D has a stronger interaction with the helium film, and the adsorption energy increases over 3~K on $^4$He~\cite{Mosk2001}. To our knowledge, there are no experimental values for the adsorption energy of T. Following the trend between H and D and theoretical estimates, one may expect that it lies between 4 and 5~K for $^4$He and 1.5-2.5~K for films of $^3$He~\cite{Stwalley1982, Goodman1989}. This would limit the lowest experimental temperatures to around 300~mK for D and 500~mK for T, provided that the recombination rate constant remained the same, which is actually not the case, as we shall see in below. 

Increasing temperature above 0.7~K leads to a high density of He vapor, which enhances three-body recombination in the gas phase, thus setting an upper limit of operating temperatures. Another effect which may occur at high enough temperature is the solvation of H, D or T in the bulk of helium liquid which leads to a sticking of the atoms to a solid substrate beneath the film and subsequent recombination. This process limits the highest temperatures at around 0.1 of the solvation energy $E_s$. For H, D, and T the predicted solvation energies are $E_s \sim 36~\text{K}$, $\sim 14~\text{K}$, and $\sim 6-7~\text{K}$~\cite{Kurten1985, Saarela1993}, accordingly. These data limit the accessible range of temperatures in experiments at $\sim0.1-0.7$~K for H, $\sim0.25-0.7$~K for D, and $\sim0.5-0.7$~K for T. Note, that for T the recombination in the helium vapor and solvation in the film set nearly the same 0.7 K upper limit of operating temperature.

Some chance remains that $^3$He-$^4$He mixture films reduce $E_a$ for tritium by a factor of $\sim3$, as observed for hydrogen, leaving the possibility of experiments at lower temperatures, down to 0.2 K. However, due to the very close values of masses of T and $^3$He, the solvation energy may strongly decrease so that T atoms will not occupy physisorbed state at the surface of $^3$He film, but dissolve in its bulk at much lower temperature than for $^4$He.  

The two-body recombination on the surface of helium films is the main loss mechanism in experiments with the high-field-seeking atomic states, which are allowed to collide with the walls. The rate of this process is characterized by a two-dimensional cross-length $l_{\mathrm{rec}}$, analog of the 3D cross-section for inelastic processes. Few experiments that attempted to work with D reported much faster recombination rates than those for H with maximum-obtained bulk densities at least an order of magnitude lower. These high rates could not be explained by the factor of 2.5-3 increase in the adsorption energy (see Table~\ref{tab:adsorption_solvation}). Discussions in refs.~\cite{Arai1998, Mosk2001} considered a complicated behavior of the effective recombination cross-length dependent on the magnetic field and temperature, as well as the possibility of resonant recombination at certain values of ~\textit{B} (Feshbach resonances). For the case of deuterium, the recombination cross-length can exceed the value for H by 2-3 orders of magnitude for a certain magnetic field and temperature range. For T, the only experimental attempt ~\cite{Tjukanov1988} did not succeed in stabilizing the T gas below 1~K. Solvation in the helium film and a large recombination cross-section were proposed for the possible explanation. In conclusion, the superfluid helium film coverage which successfully prevents surface recombination in the H gas, works poorly for D and most likely will not work for T~\cite{Reynolds1991, Tjukanov1988}.

\subsection{Scattering of H isotopes on each other and on He}
Elastic collisions that do not change the spin state of colliding atoms lead to the recovery of thermal equilibrium and are essential for evaporative cooling. Inelastic rates typically lead to a change of the hyperfine state which lead to the loss of the atoms from a magnetic trap as discussed in the next section. Scattering of atoms with helium vapor is important for the transport of hydrogen isotopes into the trapping cell and defines the efficiency of buffer gas cooling, which is one of the key effects in this proposal.

In the zero temperature limit, the elastic collision cross-section $\sigma_{el}$ is related to the $s$-wave scattering length $a_s$ as $\sigma_{el}=g_2 4\pi a_s^{2}$, where $g_2$ is the two-body correlator defined by quantum statistics and identical properties of colliding particles. $g_2=$ 1, 2, 0 for two distinguishable particles, for two indistinguishable bosons, and for two identical fermions accordingly.

The value and sign of the scattering length depend on the details of the interaction potential. The triplet potential has a shallow well arising from the van der Waals attraction and, under certain conditions, may support weakly bound $s$-wave states, or dimers leading to an enhancement of the scattering length. Neglecting mass-dependent corrections, the shape of the triplet potential for spin-polarized hydrogen isotope pairs is the same. Due to the absence of electronic spin of He isotopes in the ground electronic state, scattering processes are purely elastic. The reduced mass $\mu$ of the atomic pair is an important parameter governing the scattering properties. At a scattering resonance (or $s$-wave resonance), the scattering length diverges and changes sign, passing from $-\infty$ to $+\infty$. For the H-H pair potential, the bound state occurs at a reduced mass of $\sim 4\,\mu_{\mathrm{H-H}}$~\cite{Jones2026}, while for H-He it appears at $\sim 3.3\,\mu_{\mathrm{H-H}}$~\cite{Elliott2025}, where $\mu_{\mathrm{H-H}}$ is the reduced mass of the H-H pair. The scattering length and collisional cross-section then depend on how far the reduced mass is from the resonance value. Scattering lengths and elastic cross-sections for various pairs of H and He isotopes in the zero-temperature limit have been calculated by several authors~\cite{Jamieson1995, Stwalley1982, Stwalley2004, Jones2026}; a summary is presented in Table~\ref{cross lengths}. Of particular significance for this work are the very large collision cross-sections for the heaviest pairs considered, and especially for T-T and T-He collisions~\cite{Elliott2025, Jones2026}.

As noted above, neither the H--H nor the H--He pair potential supports an $s$-wave bound state for the masses of the naturally occurring isotopes. In contrast, the He--He pair potential does support such a state for the $^4$He--$^4$He pair, leading to an enhancement of the scattering length by an order of magnitude compared to the $^3$He--$^4$He pair which has slightly smaller reduced mass~(see Table~\ref{cross lengths}).

\begin{table*}[ht]
   \centering
    \begin{tabular}{||l||c|c|c|c|c|c|c|c||}
        \hline \hline
                                          &  D-D~\cite{Stwalley2004,Jamieson1995}    & T-T~\cite{Elliott2025} & H-T~\cite{Stwalley2004} & T-$^3$He~\cite{Jones2026} & T-$^4$He~\cite{Jones2026} & $^3$He-$^3$He~\cite{Stwalley2004,Jamieson1995} & $^3$He-$^4$He~\cite{Stwalley2004,Jamieson1995} & $^4$He-$^4$He~\cite{Stwalley2004,Jamieson1995} \\
        \hline \hline
        $a_s$, \AA                        & -4  & -42                  & -0.85                    & -8.8                     &  -22                      & -7   & -17   & -110\\
        \hline
        $\sigma_{el}, 100 \cdot$\AA$^{2}$  &  2   & 450                  & 0.09                     &  9.7                    & 61                      & 6.4       & 36     & 3200\\
        \hline \hline
    \end{tabular}
    \caption{$S$-wave scattering length and elastic collision cross-sections for different hydrogen isotopes and helium atoms in the zero temperature limit. For the fermion pairs we consider the case of distinguishable atoms, e.g. in different hyperfine states. Data for the D-D and helium pairs were calculated in refs.~\cite{Stwalley2004} and \cite{Jamieson1995} using somewhat different interaction potentials. The difference for the $a_s$ does not exceed 10$\%$ and we present average values. For the temperature dependent cross sections of hydrogen and helium isotope pairs see ref.~\cite{Jones2026}.}  
    \label{cross lengths}
\end{table*} 

The trappable low-field seeking states \textit{c} and \textit{d} are higher in energy than two other hyperfine states \textit{a} and \textit{b} (see Fig.~\ref{HF diagram}). Relaxation to the lower energy states is threshold-less and does not vanish even at zero temperature. Spin-exchange during two-body collisions is the fastest loss channel. It occurs in collisions where the "mixed" state \textit{c} takes part, leading to a rapid depletion of the \textit{c} state in the trap. The remaining "pure" and doubly polarized state \textit{d} is much more stable since spin-exchange does not work in this case. However, the dipolar interaction during collision of two \textit{d} atoms may lead to spin flip and relaxation to untrappable states. This channel of dipolar relaxation is the main loss mechanism that limits the lifetime and maximum densities of the trapped gas. This is a second order process with the characteristic decay time inversely proportional to the gas density. For H, the two-body dipolar relaxation rate constant is $G_{dd}\sim10^{-15}$~cm$^3$/s~\cite{Stoof1988}, which implies a lifetime of $\sim10^3$~s at the density $n=10^{12}$~cm$^{-3}$.
 
Stability of the trapped T gas is basically determined by relaxation processes similar to those in H. Both atoms are bosons and have a similar hyperfine structure. However, the aforementioned enhancement of the collisional cross-sections for larger mass~\cite{Elliott2025} has a strong effect. For the \textit{d}-\textit{d} relaxation of two T atoms in a magnetic field of 1~mT and temperature of 1~mK, typical for H trapping, the increase reaches a factor of $\approx50$ compared to that for H-H collisions. The dipolar rate for T-T collisions has a much stronger temperature dependence, and the difference with H can be decreased down to a factor of $\approx5$ by increasing temperature to \textit{T}~=~100~mK.
Faster dipolar relaxation leads to an increase in the loss rate and a decrease in the lifetime of the trapped T gas by the same factors at equal density. To our knowledge, magnetic trapping of T$\uparrow$ has not been attempted to date, and the theoretical predictions above are still waiting for experimental confirmation.

The situation is very different for fermionic D$\uparrow$. Due to the Pauli principle, identical D atoms in the same hyperfine state avoid approaching each other, and $s$-wave scattering is forbidden. For D, theory predicts the dipolar relaxation decreasing linearly with temperature as $G_{\zeta\zeta}\approx10^{-14}\cdot T$ cm$^3$/s~K~\cite{Koelman1988}. Remarkably, the doubly polarized D gas is becoming more stable at lower temperatures, and its decay time at 1~mK is two orders of magnitude larger than that for H, while the thermalization rate remains fairly large~\cite{Koelman1988}. Study of magnetically trapped D at ultra-low temperatures has a great potential for precision spectroscopy and will be attempted in the nearest future after improvements of the atomic source suggested in this work. 

\subsection{Stabilization of H, D, and T in solid molecular matrices}\label{solidMatrices}
Matrix isolation of unstable radicals and atoms is a well-established method in experimental physical chemistry. Unpaired atoms are stabilized in solid inert crystals of H$_2$, He, Ne, or Ar. The main fundamental interest in studies of H and its isotopes in such matrices is related to the possibility of quantum diffusion at ultra-low temperatures, the possibility of observation of superfluidity/supersolidity of impurity atoms.

Several methods are known for producing matrices with embedded atoms. We briefly describe two methods relevant to this work: dissociation with a cryogenic RF discharge and with electrons resulting from the $\beta$-decay of T$_2$. The first technique is based on an \textit{in-situ} dissociation of the matrix molecules using pulsed RF discharge in the helium vapor above solid films of hydrogen below 1~K introduced by~\citet{Ahokas2009}. The molecules are split by the impact of the electrons of the discharge having energies of the order of 100~eV. It turned out that a fraction of the atoms is evaporated during the RF pulse, and the H and D atoms in a gas phase can be accumulated in the sample cell when its surfaces are covered by a superfluid helium film. The operation of the cryogenic dissociator which we will describe in the following is based on this effect.

The second dissociation method is based on the natural radioactivity of T, which is mixed into solid films~\cite{Sharnoff_1963,Collins1992}.
%Collins~\textit{et. al.}~\cite{Collins1992} first observed that large atomic concentrations accumulate in solid film containing T$_2$ at low temperatures.
The dissociation of molecules is caused by the $\beta$-decay electrons generated inside the films. Even a few-percent concentration of T$_2$ or HT in the sample is sufficient to produce samples with large ($\gtrsim10^{20}$ cm$^{-3}$) concentrations of H and D atoms in various matrices~\cite{Sheludiakov_thesis}.

A detailed study of tritiated solid hydrogen films below 1~K was performed at the University of Turku with direct atom diagnostics using the magnetic resonance methods: ESR and ENDOR~\cite{Sheludiakov_thesis, Sheludiakov2017}. T$_2$ films of 35 and 250~nm thickness were deposited onto the surface of a quartz microbalance~(QM) which provided accurate control of the film thickness during deposition and measurements. The QM gold electrode also served as a mirror of the Fabry-Perot resonator connected to an ESR spectrometer operating at 128~GHz. For a film 250~nm thick, ESR lines of T were detected a few minutes after deposition and grew rapidly, reaching a maximum concentration of $\approx2\cdot10^{20}$~cm$^{-3}$ ($\approx 0.5\%$ relative to the density of molecules) after three hours. Then, depending on the storage temperature, periodic heat spikes were observed in the sample cell followed by a 30-40~\% decrease in the density of T and $\sim5 \%$ decrease in the thickness of the film. This behavior was explained by an explosive recombination of the atoms in the films after they reached some critical density. The explosions were not observed in a thinner T$_2$ film of 35~nm thickness and were suppressed by condensing a superfluid helium film on top of the 250~nm T$_2$ film. The maximum T densities reached during storage decreased by a factor of 1.8 after heating the 250~nm film from 0.1 to 1~K, and the time between explosions has increased by a factor of $\sim2$. Clearly, thermal explosions can be avoided by improving the heat removal from the T$_2$ film or if the atoms can be removed from the solid matrix (e.g. by running the RF discharge) before a large concentration is accumulated.

\section{Cryogenic dissociator}
\subsection{Hydrogen and \textit{hfs} of deuterium}
The cryogenic hydrogen dissociator of H operating below 1~K was first realized in the group of W.~Hardy at the University of British Columbia~\cite{Jochemsen1982, Statt1985}. The method is based on a cryogenic discharge in a helium vapor and is similar to the technique mentioned above and used to produce H atoms inside solid molecular H$_2$ matrices. The molecules in the solid are split by impacts of the electrons from a plasma discharge produced during short pulses. Some of the atoms evaporate into the volume of the dissociator chamber and are then pushed by magnetic field gradients out of the dissociator. This technique served as a source of low field seeking H atoms for loading magnetic traps. Atoms have been transported to the trap in two ways: by
loading hot atoms through a short tube for subsequent thermalization and isolation from the walls of the trap ~\cite{Hess1987, Doyle1991, vanRoijen1988}, or by pre-cooling the atoms to $\sim100$~mK in several stages of thermalization before they enter the trap. The latter technique was described in a previous publication of the Turku group~\cite{semakin2025}, where the operation of the cryogenic dissociator for H was described in detail. Atomic fluxes close to $10^{14}$~s$^{-1}$ of \textit{lfs} H have been obtained by this technique and were successfully used to load a magnetic trap with a 3~L physical volume~\cite{semakin2025}.

It is useful to consider what happens with the helium film and its vapor in the cryogenic dissociator during and after the RF discharge pulse. In a typical operation with H~\cite{semakin2025}, the discharge is driven by RF pulses of $\sim1$~ms length followed by $\sim20$~ms off time. The RF power is adjusted to the maximum value, which the dilution refrigerator can tolerate but is lower than the threshold for full evaporation of the helium film inside the dissociator chamber. The energy released in the dissociator resonator during the 1~ms RF pulse is sufficient to evaporate $\approx3\cdot10^{-7}$ moles of $^{4}$He which has a latent heat of evaporation of $\approx70$~J/mole at 0.6~K~\cite{RusselDonnelly}. This is nearly half of the total amount of helium that covers the walls of the dissociator. During operation, the dissociator's average temperature increases from 0.6 to $\approx0.64$~K, leading to an increase in the $^4$He vapor density in the chamber from $4.5\cdot10^{15}$ cm$^{-3}$ to $1.05\cdot10^{16}$~cm$^{-3}$~\cite{Dijk1960, Kaufman1993}. We used here the average temperatures during pulsed operation measured by a thermometer at the dissociator body. Obviously, the temperature during the pulse is somewhat higher, and therefore the estimate above represents a lower bound for the density of helium vapor during the pulse. 

We can estimate the upper limit if we assume that all of the liquid evaporated during the pulse is instantly converted into vapor filling the dissociator volume of 13~cm$^{-3}$. This gives a vapor density of $1.4\cdot10^{16}$~cm$^{-3}$, quite close to the lower bound estimate. Close match of these bounds indicates that the vapor temperature is close to the dissociator's body temperature of 0.64 K. 

Helium vapor during and after the pulse flows out of the dissociator and is re-condensed in the colder regions of the transfer line. The vapor is dense enough to entrain D(T) atoms that were produced during the discharge pulse, and it pushes them towards the colder regions of the transfer line. This effect, similar to the operation of a diffusion pump, is useful to increase the efficiency of the atomic source. It was initially observed in the first experiments on the stabilization of H$\downarrow$~\cite{Silvera1980} and was given the acronym HEVAC (Helium Vapor Compressor). The superfluid flow along the walls of the transfer line returns helium into the dissociator chamber. Such circulation of helium exists in any system where there is a gradient of temperature and often creates problems because of the associated heat load to the lower temperature parts of the system. 

Buffer gas cooling with the helium vapor was used for trapping various molecular species in a dipole magnetic trap by a J.~Doyle's group \cite{Hutzler2012, Michniak2004}. They used helium buffer gas for pre-cooling various molecules to trappable energies using the same strength of magnetic potential and range of the buffer gas densities discussed here. After cooling, the buffer gas was pumped out and this has to be done slow enough for not removing the trapped species together with the buffer gas. Simulations of the pumping process in the gradient of magnetic field revealed that fast pumping, on the time scale of several ms creates strong enough "wind" to blow out the trapped gas. This undesirable process for their experiments, is exactly what happens in our dissociator shortly after the 1~ms RF pulse: the wind of helium gas moves entrained T atoms into the atom guide and transfer them further to the colder regions. Similar effects were utilized in experiments of the C.~Cesar's group at UFRJ where the Ne or H$_2$ was used as a buffer gas and various ions were entrained in it after sublimation from solid matrix~\cite{Azevedo:2023kgf}.  

A cryogenic dissociator of this type was also used in experiments with atomic \textit{hfs} of D, for a hyperfine resonance study of an unpolarized gas in a zero magnetic field~\cite{Reynolds1991, Hayden1995} and with two-dimensional D in a strong magnetic field~\cite{Mosk2001}. The fluxes were a factor of 5-10 smaller than for H. The large adsorption energy on the helium film and fast recombination rate on the surface of the dissociator and trapping cell were proposed as a possible explanation.

\subsection{Deuterium and tritium \textit{lfs} }
Attempts to load the \textit{lfs} states of atomic deuterium to a magnetic trap using same dissociator and loading technique as for H were performed at MIT and were unsuccessful~\cite{Steinberger2004}. Some flux of atoms was detected entering the trap, but atoms quickly recombined after the dissociator was turned off. At this stage the trapped gas was still rather warm $\sim 300$~mK, could not be magnetically isolated from the walls and easily adsorbed on the superfluid film surface. Another reason may lie in the transfer line between the dissociator and the trapping cell, which was very short. 

\begin{figure*}
    \includegraphics[width=\textwidth]{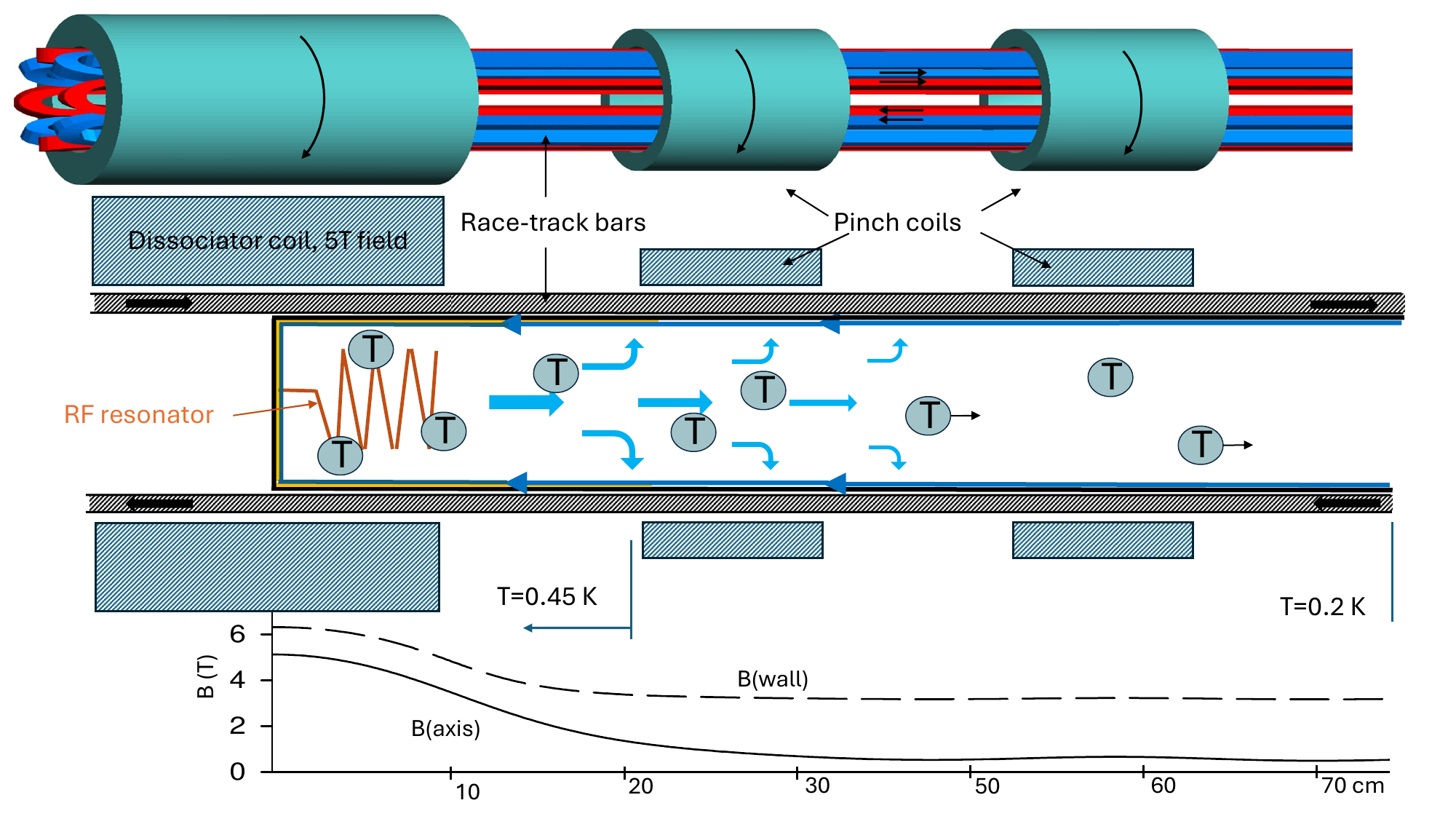}
    \caption{Schematic of the cryogenic source of D and T. 3D sketch of the magnet system is shown at the top. Cross-section of the dissociator and the atom guide is shown in the middle, and the profiles of the of the magnetic field on the wall and on the axis are shown at the bottom. Overall length of the source is slightly above 1 m, last 25 cm are not shown. The solid molecular D$_2$(T$_2$) layer at the tube inner walls in the left is shown with the yellow line and the superfluid helium film lining inner surface of the tube is shown with the blue line. Solid film of D(T) is condensed using a thermally isolated and heated above 10-15 K capillary (not shown).}
    \label{T source}
\end{figure*}

Our proposal for a cryogenic source of \textit{lfs} D and T is based on the dissociation technique below 1~K described above for H. However, further cooling of atoms and their transport to the magnetic trap cannot rely on thermalization of atoms by collisions with the superfluid-helium-covered wall of the transfer line. As mentioned above, the latter may work somewhat with \textit{hfs} of D, may not work well with \textit{lfs} of D, and most probably will not work with T. However, the presence of helium cannot be avoided in the dissociator chamber since helium vapor is necessary for running the discharge in the RF resonator and for having the HEVAC effect which provides thermalization of atoms and pushes them out of the dissociator. We consider a similar construction for the D(T) dissociator as utilized in Turku for H, combined with a radial confinement by multipole system of linear conductors; see Fig.~\ref{T source}.

Aiming on higher fluxes of atoms we scale up the size of the dissociating region by a factor of $\sim5$ and increase the RF power by an order of magnitude, running discharge with 10~mW on average.  We consider a cylindrical channel of 10~cm diameter, $\gtrsim100$~cm long.  The left closed end is cooled to $\approx0.45$~K by a $^3$He refrigerator (not shown). It forms a dissociation region, and the tube continuing to the right, guides the atoms into the magnetic trap. Dissociation is performed in the RF discharge in the helical resonator with a resonance frequency 100-500~MHz, operating in a pulsed mode. The dissociator tube is located inside a multi(4-8)-pole system of race-track type coils which create a 3~T magnetic field at the tube wall. A superconductive solenoid with 5~T central field is mounted on top of the race-track bars at the left end of the dissociator tube. Extra pinch coils located to the right from the main dissociator coil, allow control of the axial field profile. The field in the right end of the atom guide cannot be lower than that at the bottom of magnetic trap. Otherwise, the atoms will be reflected from the trap and accumulate in the atom guide. Right end of the atom guide is thermally anchored to a mixing chamber of a dilution refrigerator and cooled to $\sim$0.1~K. Solid molecular T$_2$ is condensed onto the tube inner walls at the left side. This is done by spraying it from a thermally isolated and heated to  15-20 K capillary. Then, the tube inner surface is lined with a superfluid helium film.

\begin{figure}
    \includegraphics[width=\linewidth]{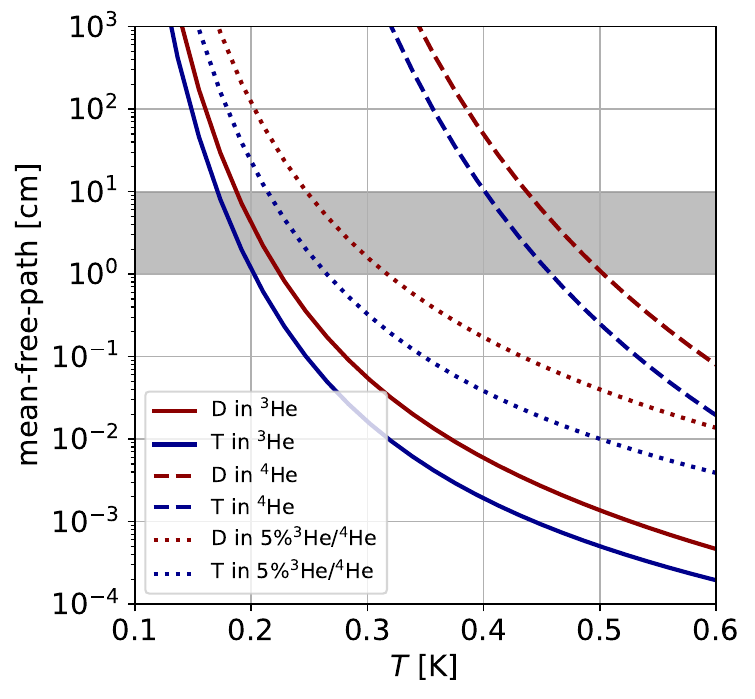}
    \caption{Mean-free-path of deuterium and tritium atoms in saturated $^{3,4}$He~(solid, dashed) and 5~\% mixture~(dotted) $^3$He in $^4$He vapors as a function of temperature above the liquid surface. Curves are based on saturated vapor density data from~\cite{Huang2006, NIST4HE80} and the energy-dependent scattering cross-section~\cite{Jones2026}. Horizontal gray band denotes the 1-10 cm region, the dimensions of the dissociator and the transfer line where the buffer gas cooling occurs.}
    \label{Mean free path}
\end{figure}

Lowering temperature leads to a very steep decrease in the saturated vapor density of helium in the considered temperature range. Mean-free path $\lambda$ between collisions of hydrogen and helium isotopes increases, and may become larger than the dimensions of the chamber where we utilize buffer gas cooling, thus limiting the operating temperature from below. In Fig.~\ref{Mean free path}, we present the dependence of the mean free path for D and T in saturated vapor of $^4$He and $^3$He as a function of temperature based on the data of refs.~\cite{NIST4HE80, Huang2006, Jones2026}. We find that for the dimensions of the dissociation region and the atoms transfer line (10 cm), buffer gas cooling will cease to work at $\sim450(400)$~mK for D(T) atoms in $^4$He vapor and at $\sim180(170)$~mK in $^3$He vapor. 

For effective isolation of the \textit{lfs} atoms from the wall, their thermal energies must be 5-10 times lower than the height of the potential barrier. Looking at the magnetic field profiles along the axis of the atom guide, we can see that the radial magnetic barrier is $\sim 1$~T in the end of dissociation region, increases to $\sim2$ T at the distance of $\approx20$ cm on the right and reaches $\approx3$ T in the middle of the source (right end of the atom guide shown in Fig.\ref{T source}). It is important that the atoms pass quickly the first region with minimal number of collisions with the walls. This a region where the HEVAC has strongest effect and helps to reach the condition above. Using simple model of the vapor expansion and diffusion, in Appendix A, we analyze the helium vapor and D(T) gas behavior in the dissociator region and the transition section just after it. We find that the helium vapor density due to the condensation on the wall decreases on the time scale of 1~ms, and the vapor spreads in a laminar viscous flow over the distance of several diameters. D(T) gas entrained in the vapor flow, diffuses towards the wall with order of magnitude smaller rate, which justifies feasibility of the proposed source design.

We note that the main parameters of the RF discharge: the pulse width, repetition rate and RF power may be varied in a fairly wide range for optimization of the flux. In the smaller version of H dissociator, we were able to get reasonably large fluxes varying them by an order of magnitude provided the average power was kept the same. However, in the design suggested in Fig.~\ref{T source}, the vapor density after the RF pulse, characteristic condensation and diffusion rates may have stronger influence on the output flux and should be adjusted via changes in the above mentioned discharge parameters. This should be verified in experiments.

Pure $^3$He cannot be used to line inner surface of the atom guide because it is not superfluid in the considered temperature range. We can use $^3$He-$^4$He mixture films instead. Mixtures up to $\sim 5\%$ of $^3$He were previously used by the Turku cryogenic dissociator of \textit{hfs} of H~\cite{Safonov2001}. The vapor density above such mixture in equilibrium will be reduced roughly proportional to the concentration of $^3$He~\cite{Nacher1994}. In case of the temperature gradients in the transfer line and due to the HEVAC effect, density will be higher than in equilibrium. Even for the worst case of the vapor density being $5\%$ from that above pure $^3$He, we can easily compensate such density reduction by raising the temperature by 30-50~mK (see Fig.~\ref{Mean free path}), thanks to the very steep dependence of the vapor density on temperature. Working with D will also require slightly higher temperatures than for T due to a smaller collisional cross-sections~\cite{Jones2026}, as can be seen in Fig.~\ref{Mean free path}. For the 5$\%$ $^3$He-$^4$He mixture film, the buffer gas cooling will cease to work at 200-250~mK, slightly higher than for pure $^3$He. 

In the considered in the Fig.\ref{T source} example, we presented the total length of the atom guide of about 1 m, ending at temperature of 100 mK. This choice of the length is mainly dictated by the necessary temperature gradient in the atom guide part after the dissociator region. The section between 0.45 and 0.2~K should be sufficiently long to ensure several collisions of D(T) gas with helium in its lowest temperature part where the gas density starts to be at the limit of the buffer cooling. We suggest that this region should have the length exceeding 2-3 tube diameters. One may think of further cooling below 0.2~K adding extra stages to the atom guide and using dynamic evaporative cooling, as proposed by the Project 8~\cite{P8_MECB_2025}. Design of these parts depend on the desired temperature of the atoms and final coupling to the magnetic trap. We do not consider this issues here.

\subsection{Self-dissociation of T$_2$ in the dissociator chamber by $\beta$-decay electrons.}
So far, we have considered that the dissociation of molecules in the solid T$_2$ layer in the dissociator chamber as driven by by the electrons of the RF discharge. As described in Section~\ref{solidMatrices}, electrons resulting from the $\beta$-decay of T also effectively produce atoms in the solid layer. In the previous work of the Turku group with T in T$_2$~\cite{Sheludiakov2017}, very thin films of a maximum of 250~nm were used, limited by the exemption limit for work with radioactive materials (1~GBq). We found that each $\beta$-electron dissociated about 25 molecules and produced 50 atoms. With a dissociation energy of 4.6~eV, this corresponds to a production efficiency of $\lesssim0.05 \%$. The penetration depth of 5.7~keV electrons in solid hydrogen is about $\sim3.5 ~\mu$m~\cite{Schou1978}, and we expect that the dissociation number per T decay will be substantially larger for thicker films. 

Let's consider the self-dissociation effects in the proposed above source. The dissociation region will be covered inside by a T$_2$ layer of 1~$\mu$m thickness. For the 10~cm diameter tube of 10 cm length, the total amount of T$_2$ would be $\approx3\cdot10^{-3}$ moles and the rate of $\beta$-decay events in this layer would be $\approx 6\cdot 10^{12}$~s$^{-1}$. Taking the same dissociation efficiency of 50 atoms per T decay, we find a lower bound for the production rate of atoms in solid T$_2$ as \.N$_T\approx 5\cdot10^{14}$~s$^{-1}$. If we assume that the efficiency scales linearly with thickness, then it should increase to 200~events/decay, and the T production rate should reach $\sim2\cdot10^{15}$~s$^{-1}$. This is larger than the T flux we may get from the dissociator by running the pulsed RF discharge above the solid film with 10~mW power. The total heat released by the decays of T in the estimate above is evaluated as $\approx10$~mW. The self-dissociation of T$_2$ seems to be a very efficient mechanism and may provide better efficiency than the RF discharge. As we mentioned in Section II E, the presence of the superfluid helium film improves the cooling of the T$_2$ films and helps to avoid thermal explosions and partial evaporation of the film. One may try to increase the thickness of T$_2$ further if the cooling by the helium film will still be able to stop thermal explosions. Finally, both dissociation methods -- running the RF discharge on top of the T$_2$ film, and self-dissociation -- will work together, and reaching the flux of atoms above $2\cdot10^{15}$ s$^{-1}$ seems quite realistic. 

\subsection{Practical implementation}
The KATRIN++ and Project 8 collaborations plan to use large magnetic traps to store T atoms for neutrino mass measurements. The next generation experiments needs at least the activity of KATRIN or higher, thus a N$_T$ of $10^{19}$ or higher is required. Let's evaluate volume $V$ of the trap, gas density $n_T$ and atomic flux $\Phi_T$ required for reaching this target. We consider the dipolar relaxation as the main loss mechanism and that the incoming flux of atoms in the trap compensates for this loss rate: $\frac{dN_T}{dt}=-G_{dd}N_T^2/V=\Phi_T$. Then, we get for the product of the flux and trap volume: $\Phi_T\cdot V=N_T^2\cdot G_{dd}=2\cdot10^{23}$ cm$^3$/s. Here we used the value for the loss rate constant $G_{dd}\approx2\cdot10^{15}$ cm$^3$/s ~\cite{Elliott2025} for realistic experimental conditions of $T~=~100$~mK and $B~=~1$~mT. This means that with the atomic flux of $\Phi_T=10^{15}$ s$^{-1}$ expected from the proposed here source, a trap volume of $\sim200$ m$^3$ would be required. Project 8 previously studied a possibility of using magnetic trap of the ten-plus cubic meters size~\cite{Esfahani_2017}, but even this number looks challenging. However, if the target requirement can be reduced by an order of magnitude, then the volume of 2 m$^3$ would be sufficient, thanks to the second order dependence of the relaxation rate on density. The gas density in such trap would be $n_T=5\cdot10^{11}$ cm$^{-3}$. 

In principle, the temperature range of 0.22-0.44~K expected for the T source described above is already sufficiently low for magnetic trapping. A large ($V\sim 8~l$) superconducting trap with 3.1~T (2~K) magnetic barrier has been built in the High Energy Accelerator Research Organization (KEK) Laboratory in Japan~\cite{KekCryogenics2014} and used to trap ultra-cold neutrons~\cite{KekNIM2009}. However, for the large trap volumes considered, smaller fields and lower temperatures are preferable. Also, as we discussed in Section II~B, the minimal atom loss due to dipolar relaxation is reached at around 0.1~K. In order to increase the efficiency of the buffer gas cooling even once the mean-free path of T-He collisions becomes larger than the transfer line diameter, slowing down the flow of the T beam along the line can be utilized. This can be done by introducing distortions or bends in the atom guide as suggested by the Project 8~\cite{Esfahani2025, LindmanThesis} and QTNM~\cite{Amad_2025} projects. Another option to cool the D(T) gas flowing in the transfer line without buffer gas was recently proposed by the Project 8 collaboration~\cite{P8_MECB_2025, LindmanThesis}. They propose using evaporative cooling that allows the high-temperature tail of the energy distribution of the flowing gas to overcome the radial magnetic barrier and be removed.

For experiments aiming at precision spectroscopy and the observation of the gravitational quantum states of D or T, we propose feeding the atomic beam out of the cryostat with the dilution refrigerator to room temperature instruments, so that the experiments can be done using the same techniques as is done with a 6~K nozzle~\cite{Matveev2013, Killian2024}. Since the superfluid helium film is supposed to line the inner walls of the transfer line, extending it to room temperature requires suppression of the film and preferably also He vapor flow together with the atomic beam. This can be done by installing a superfluid film cutter based on the evaporation and condensation of the film in a specially designed geometry~\cite{Kaufman1993, Ishikawa2010} or by suppressing superfluid flow on a surface coated with Cs~\cite{AnthonyPetersen2024}.

For experiments with slow beams, we may employ the pulsed mode of operation of the dissociator.
The average flux of $10^{15}$~atoms/s considered above consists of pulses 1~ms long followed by a delay of 20~ms. The number of atoms in each pulse is $\approx2\cdot10^{13}$ and close to that reported by Helffrich~\cite{Helffrich1987} for a similar type dissociator for H.
The pulsed D(T) beams at 0.2-0.4~K can be slowed at room temperature using Zeeman decelerators. This technique is based on applying decelerating magnetic field pulses synchronized with the motion of the bunches of atoms in the beam. For H, a device consisting of 12 deceleration stages decreased atom velocities from 520 to 100~m/s~\cite{Merkt2008} with an average deceleration of 35~m/s per stage. Two to three such stages will be enough to completely stop the 0.2-0.4~K cold D (T) atoms exiting from the source proposed in this work.

Practically, cooling the dissociator to 0.2~K can be done by thermally anchoring its chamber to the cold plate of the dilution refrigerator, typically operating at 0.1-0.2~K. Working with pure $^4$He requires higher temperatures of around $0.4$~K, which can be obtained by using a $^3$He refrigerator and cooling the next stages of the transfer line with a dilution refrigerator. In the Turku lab, a cryogenic system with two refrigerators in one cryostat - dilution and $^3$He type - was used for H \textit{hfs} studies. The $^3$He refrigerator also has substantially larger cooling power at $\sim$0.4~K than the dilution refrigerator cold plate.

Our strategy to increase the atom flux from the source is to increase the discharge RF power and dimensions of the dissociator chamber. As a reference, the H source in Turku runs at an average power of 1~mW as reported previously~\cite{semakin2025}. The flux of atoms saturated at around this power value, and increasing it further by a factor of 2 stopped the flux completely. As a possible explanation, we consider an insufficient backflow of superfluid film into the dissociator chamber that could not be compensated for by evaporation of helium inside, leading to the helium film in the dissociator drying out at RF powers exceeding some critical value of $\sim2\,$mW. This problem is solved in our proposal here by increasing the diameter of the transfer line by the factor of 20 and lowering the operating temperatures to 0.45-0.5 K. The latter strongly reduces the heat of refluxing helium vapor. In the small H dissociator in Turku, the $^4$He vapor re-condensing from the dissociator on the first accommodator at $\sim0.25$ K delivered $\approx 0.4$ mW of heat. In the proposed here source this power is reduced below 0.1 mW. One may think of increasing the RF power further. Then, the maximum dissipated power in the dissociator will be limited by the available cooling power at 0.2 or 0.4~K. In the latter case, using a $^3$He refrigerator, it can be $\gtrsim50$~mW, which may potentially increase the atomic flux by a factor of 5 and reach values $\sim5\cdot10^{15}$~s$^{-1}$. 

Using $^3$He-$^4$He films may be challenging when running the dissociator at the highest power of 10~mW. It is known that the transport properties of such films and the critical superflow velocity are reduced due to the presence of $^3$He, which contributes to a normal component of the film~\cite{Hallock1995}. In this case, the concentration of $^3$He can be reduced to a level required for stable operation of the dissociator; the operation may resume with pure $^4$He films. 

\section{Conclusions}
We propose a concept of a cryogenic source of atomic tritium based on a pulsed RF discharge below 1~K, a technique successfully used to produce large fluxes of atomic hydrogen. 
We expect an extra enhancement of the dissociation rate due to contributions of electrons resulting from the $\beta$-decay of T.
Working with T requires 
to prevent the interaction of atoms with the walls
where they adsorb and recombine, even if a superfluid helium film is used. We suggest using a buffer gas cooling technique to provide a thermal link between atoms and the walls without physical contact. The atoms are expelled from the walls by magnetic field gradients. 
Thermal contact via the buffer gas may be easily adjusted by small changes in the temperature distribution in the gas transfer line. Optimal geometry and operating parameters may be found in experimental tests of a source prototype. We evaluate that the fluxes of T atoms exceeding $10^{15}$~s$^{-1}$ can be reached at temperatures of $\sim200$~mK, optimal for magnetic trapping and other experiments with the beams of slow D and T atoms.

\section{Acknowledgments}
This project was supported by the Jenny and Antti Wihuri foundation. FN and PY acknowledge financial support from CNRS IRP GRASIAN 2024-2028 and ANR (grant ANR-23-CE30-0033-01). The authors thank Ben Jones and Alec Lindman for discussions and commenting the manuscript.

\begin{appendices}
\section{Diffusion in the gas flow after the RF pulse}
Let's analyze what happens during the operation cycle of the dissociator. First, we evaluate possible limits on the density and size of the $^4$He vapor cloud in the end of the RF pulse. The pulse of 200 mW power during 1 ms evaporates $\lesssim 1.8\cdot10^{18}$ $^4$He atoms. This is an upper bound, evaluated with assumption that the RF energy is consumed for evaporation with the latent heat of evaporation of $\approx$70~J/mole or $\approx$ 8~K per atom~\cite{RusselDonnelly}, and vapor atoms have the same temperature as the walls. In order to estimate the size of the cloud and density of vapor, we assume that it can expand with linear velocity close to the thermal velocity of $^{4}$He atoms, $v\approx 60$ m/s at 0.5 K. Thus, during the 1 ms pulse the cloud will spread for $L_0\sim$15 cm of the beam tube including the 10 cm region of the resonator. Then, the volume of the cloud after the pulse can be estimated as $\sim10^3$~cm$^{3}$ which gives vapor density of $n_{4}\approx1.8\cdot10^{15}$ cm$^{-3}$ in the end of the RF pulse. The vapor density estimate above is an upper limit because the helium vapor may be somewhat overheated above the wall temperature of the dissociation region, the vapor expansion is faster and the occupied volume after pulse is larger.

Analyzing helium vapor density and temperature in the small H dissociator in the Section IIIA, we found that helium vapor was overheated by $\lesssim40$ mK at 0.6 K operating temperature.  For the larger version of T(D) dissociator considered here, we assume a 10 fold larger RF power, 25 times larger surface area where the heat is removed, and 1.3 times lower operating temperature of 0.45 K. The latter increases thermal boundary resistance between liquid helium and copper surface (Kapitza resistance) by a factor of $(1.3)^3\approx2.2$~\cite{LounasmaaBook}. Taking into account all these factors we expect that the vapor overheating in the T(D) dissociator should be about the same, $\sim0.04$~K which justifies the evaluations of the vapor density and its cloud size presented above. 

Taking this estimate of the $^4$He vapor density  and collisional crossections from ref.~\cite{Jones2026}, we estimate that after the RF pulse the mean-free path between collisions of T-$^4$He is $\approx0.3$ mm, D-$^4$He - $\approx1.4$ mm, and of $^4$H-$^4$H - $\approx0.2$ mm. Since all these mean-free paths are substantially smaller than the size of the tube where the mixture of atoms and helium vapor moves, the atoms approach the wall in a diffusive motion experiencing many collisions with helium atoms which is important for their thermalization.

After the RF pulse, the vapor cloud continues to expand, its density decreases because of that and also due to condensation of helium vapor onto the walls of the atom guide. For the latter we may assume that helium atoms have very high, close to 1 sticking probability. Then, we may write an equation for the loss of the total number of atoms in the cylinder of the length $L(t)$ and diameter $d$ filled by the vapor with the density $n_4$:
\begin{equation} \label{CondLoss}
    \begin{split}
        \frac{dN}{dt}&=-\frac{1}{4}n_{4} v\left(\pi d L(t) + \pi d^2\right)\\ &= -v\left(\frac{1}{d}+\frac{1}{L_0 +\frac{v}{4}t}\right)N,    
    \end{split}
\end{equation}
where we considered linear expansion $L=L_0 + ut$ with the macroscopic expansion velocity $u=v/4$ and the cloud length just after the RF pulse $L_0$. The first term in the brackets coming from the condensation on the left end of the tube can be taken at $t=0$ for simplicity. Then, we will get exponential decrease of the atom number with the time constant $\tau_{cond}=\frac{d L_0}{v(d+L_0)}\approx 1$ ms for our case of $L_0 \approx d$.  The characteristic length increase of the cloud after the time $\tau_{cond}$ is approximately equal to the cylinder diameter of 10 cm. This means that the excess of helium vapor created by the RF pulse will disappear at the distance of 2-3 diameters of the atom guide and the vapor density will be close to its equilibrium value given by the temperature of the wall.

Next, we consider diffusion of helium vapor and T atoms towards the wall. Now, for simplicity, we neglect the axial expansion of the cloud and consider simple diffusion problem in axial symmetry with a diffusion constant $D\approx \lambda v/3$ which gives characteristic diffusion time for reaching the cylinder wall of radius $R$: 
\begin{equation} \label{TauDif}
\tau_{dif}=\frac{R^2}{6D}=\frac{d^2}{8\lambda v},
\end{equation}
Taking the value of 0.3 (1.4) mm for the mean-free path of T(D) in helium vapor at 0.5 K evaluated above, we find $\tau_{dif}\gtrsim 50 (10)$ ms. For both isotopes this time greatly exceeds the characteristic condensation time of helium vapor. In this estimate, we neglected effects of the magnetic field gradient and corresponding drag force on the \textit{lfs} of D(T) atoms. This will lead to even stronger suppression of the diffusion of atoms towards the wall.

Based on these simple considerations, we conclude that the T atoms will pass the 10 cm region with small radial magnetic barrier with minor loss due to the diffusion to the wall.

\end{appendices}

\bibliography{Tsource}
\end{document}